\documentclass[amsmath,aps,showpacs,prd,a4paper,10pt]{revtex4}
 \usepackage{epsf}
 \usepackage{graphicx}    % Include figure files

\begin{document}

 \newcommand{\be}[1]{\begin{equation}\label{#1}}
 \newcommand{\ee}{\end{equation}}
 \newcommand{\bea}{\begin{eqnarray}}
 \newcommand{\eea}{\end{eqnarray}}
 \def\disp{\displaystyle}

 \def\gsim{ \lower .75ex \hbox{$\sim$} \llap{\raise .27ex \hbox{$>$}} }
 \def\lsim{ \lower .75ex \hbox{$\sim$} \llap{\raise .27ex \hbox{$<$}} }

 \begin{titlepage}

 \begin{flushright}
 arXiv:0902.0129
 \end{flushright}

 \title{\Large \bf Entropy-Corrected Holographic Dark Energy}

 \author{Hao~Wei\,}
 \email[\,email address:\ ]{haowei@bit.edu.cn}
 \affiliation{Department of Physics, Beijing Institute
 of Technology, Beijing 100081, China}

 \begin{abstract}\vspace{1cm}
 \centerline{\bf ABSTRACT}\vspace{2mm}
The holographic dark energy (HDE) is now an interesting
 candidate of dark energy, which has been studied extensively
 in the literature. In the derivation of HDE, the black hole
 entropy plays an important role. In fact, the entropy-area
 relation can be modified due to loop quantum gravity or other
 reasons. With the modified entropy-area relation, we propose
 the so-called ``entropy-corrected holographic dark energy''
 (ECHDE) in the present work. We consider many aspects of
 ECHDE and find some interesting results. In addition, we
 briefly consider the so-called ``entropy-corrected agegraphic
 dark energy'' (ECADE).
 \end{abstract}


 \maketitle

 \end{titlepage}

 \renewcommand{\baselinestretch}{1.5}

%============================= section 1 ===================================

\section{Introduction}\label{sec1}
The holographic dark energy (HDE) is now an interesting
 candidate of dark energy, which has been studied extensively
 in the literature. It is proposed from the holographic
 principle~\cite{r1,r2} which is a possible window to quantum
 gravity. For a quantum gravity system, the local quantum
 field cannot contain too many degrees of freedom, otherwise
 the formation of black hole is inevitable and then the quantum
 field theory breaks down. In the thermodynamics of the black
 hole~\cite{r3,r4}, there is a maximum entropy in a box of size
 $L$, namely, the Bekenstein-Hawking entropy bound
 $S_{BH}\sim m_p^2 L^2$ which scales as the area of the box
 $A\sim L^2$, rather than the volume $V\sim L^3$. Notice that
 $m_p\equiv (8\pi G)^{-1/2}$ is the reduced Planck mass,
 whereas we use units $\hbar=c=1$ throughout. To avoid the
 breakdown of the local quantum field theory, Cohen {\it et
 al.}~\cite{r5} suggested a more restrictive bound. They
 proposed that the entropy for an effective quantum field
 theory $\sim L^3\Lambda^3$ should satisfy~\cite{r5,r6}
 \be{eq1}
 L^3\Lambda^3\,\lsim\left(S_{BH}\right)^{3/4}\sim
 m_p^{3/2}L^{3/2},
 \ee
 where $L$ is the size of a region which provides an IR
 cut-off; $\Lambda$ is the UV cut-off. That is, they replaced
 the original maximum entropy $S_{BH}$ with the more
 restrictive $\left(S_{BH}\right)^{3/4}$. In fact,
 Eq.~(\ref{eq1}) is equivalent to~\cite{r5}
 \be{eq2}
 L^3\rho_\Lambda\,\lsim\,Lm_p^2\,,
 \ee
 where $\rho_\Lambda\sim\Lambda^4$ is the energy density
 corresponding to the zero-point energy and the cut-off
 $\Lambda$. Obviously, Eq.~(\ref{eq2}) means that the total
 energy in a region of size $L$ cannot exceed the mass of a
 black hole of the same size~\cite{r5}. From Eqs.~(\ref{eq1})
 or (\ref{eq2}), it is easy to find that
 $\rho_\Lambda\,\lsim\, m_p^2L^{-2}$.

In~\cite{r6}, there is an alternative derivation of HDE by
 invoking the Bekenstein bound. For a macroscopic system in
 which self-gravitation effects can be disregarded, the
 Bekenstein bound $S_B$ is given by the product of the energy
 $E\sim\rho_\Lambda L^3$ and the linear size $L$ of the system.
 Requiring $S_B\leq S_{BH}$, namely $EL\,\lsim\,m_p^2 L^2$,
 one has the same result $\rho_\Lambda\,\lsim\, m_p^2L^{-2}$.
 We refer to~\cite{r6} for details.

In the literature, commonly the energy density of HDE is
 parameterized as
 \be{eq3}
 \rho_\Lambda=3n^2 m_p^2 L^{-2},
 \ee
 where the numerical constant $3n^2$ is introduced for
 convenience. If we choose $L$ as the size of the universe,
 for instance the Hubble horizon $H^{-1}$, the resulting
 $\rho_\Lambda$ is comparable to the observational density of
 dark energy~\cite{r7,r8}. However, Hsu~\cite{r8} pointed out
 that in this case the resulting equation-of-state
 parameter~(EoS) is equal to zero, which cannot accelerate the
 expansion of our universe. To get an accelerating universe,
 Li proposed in~\cite{r9} to choose $L$ as the future event
 horizon
 \be{eq4}
 R_h=a\int_t^\infty\frac{d\tilde{t}}{a}
 =a\int_a^\infty\frac{d\tilde{a}}{H\tilde{a}^2}\,,
 \ee
 where $H\equiv\dot{a}/a$ is the Hubble parameter; $a$ is the
 scale factor of the universe; a dot denotes the derivative
 with respect to cosmic time $t$. In this case, the EoS of HDE
 can be less than $-1/3$~\cite{r9}. For a comprehensive list of
 references concerning HDE, we refer
 to e.g.~\cite{r10,r11,r70,r72,r73} and references therein.

Obviously, in the derivation of HDE, the black hole entropy
 $S_{BH}$ plays an important role. As is well known, usually,
 $S_{BH}=A/\left(4G\right)$, where $A\sim L^2$ is the area
 of horizon. However, in the literature, this entropy-area
 relation can be modified to
 \be{eq5}
 S_{BH}=\frac{A}{4G}+\tilde{\alpha}\ln\frac{A}{4G}
 +\tilde{\beta}\,,
 \ee
 where $\tilde{\alpha}$ and $\tilde{\beta}$ are dimensionless
 constants of order unity. These corrections can appear in the
 black hole entropy in loop quantum gravity
 (LQG)~\cite{r12,r13,r14,r15,r16,r17,r18,r19,r20,r21,r22};
 they can also be due to thermal equilibrium fluctuation,
 quantum fluctuation, or mass and charge
 fluctuations~\cite{r20,r23,r24,r25,r26,r27,r28,r71,r72}. We
 refer to e.g.~\cite{r20,r29,r30} for some brief reviews. Since
 in the literature the values of the constants $\tilde{\alpha}$
 and $\tilde{\beta}$ are still in debate, we keep them as free
 parameters in the present work.

Considering the corrected entropy-area relation~(\ref{eq5}),
 and following the derivation of HDE (especially the one shown
 in~\cite{r6}), we can easily obtain the density of the
 so-called ``entropy-corrected holographic dark energy''
 (ECHDE), namely
 \be{eq6}
 \rho_\Lambda=3n^2 m_p^2 L^{-2}+
 \alpha L^{-4}\ln\left(m_p^2 L^2\right)+\beta L^{-4},
 \ee
 where $n$, $\alpha$ and $\beta$ are dimensionless constants of
 order unity. Since $n$ is given in the form of $n^2$, we only
 consider the positive $n$. Obviously, when the last two
 terms can be ignored, Eq.~(\ref{eq6}) reduces to the one of
 ordinary HDE, i.e. Eq.~(\ref{eq3}). Since the last two terms
 in Eq.~(\ref{eq6}) can be comparable to the first term only
 when $L$ is very small, the corrections make sense only at
 the early stage of the universe. When the universe becomes
 large, ECHDE reduces to the ordinary HDE.

In the present work, we will study the effects of ECHDE to the
 universe. In sections~\ref{sec2} and \ref{sec3}, we consider
 the universe which is dominated by ECHDE with $L=H^{-1}$ and
 $R_h$, respectively. We find that ECHDE can drive an inflation
 phase at the very early stage of the universe. Since the
 corrections to $\rho_\Lambda$ can be due to loop quantum
 gravity, thermal equilibrium fluctuation, quantum fluctuation,
 or mass and charge fluctuations, we consider ECHDE in both
 frameworks of loop quantum cosmology (LQC) and
 Friedmann-Robertson-Walker (FRW) cosmology. Notice that in LQC
 the Friedmann equation has been modified by the quantum
 geometry. In Sec.~\ref{sec4}, we consider the entire evolution
 history of the universe with ECHDE. In Sec.~\ref{sec5}, we
 consider the so-called ``entropy-corrected agegraphic dark
 energy'' (ECADE) whose $L$ is chosen to be the conformal time
 $\eta$ (note that we set units $\hbar=c=1$ throughout, one can
 use the terms like length and time interchangeably). In fact,
 ECADE is the entropy-corrected version of the new agegraphic
 dark energy (NADE)~\cite{r31,r32}. In some sense, NADE with
 generalized uncertainty principle (GUP) studied in~\cite{r33}
 can be regarded as a special case of ECADE. Finally, we give
 some brief remarks in Sec.~\ref{sec6}.

%============================= section 2 ===================================

\section{Inflation phase driven by ECHDE with $L=H^{-1}$}\label{sec2}
In this section, we consider the universe which is dominated by
 ECHDE with $L=H^{-1}$. Since the corrections to $\rho_\Lambda$
 can be due to loop quantum gravity, thermal equilibrium
 fluctuation, quantum fluctuation, or mass and charge
 fluctuations, we consider ECHDE in both frameworks of loop
 quantum cosmology (LQC) and Friedmann-Robertson-Walker (FRW)
 cosmology.

%============================= section 2.1 ===================================

\subsection{The case in FRW cosmology}\label{sec2a}
Here, we consider the ECHDE with $L=H^{-1}$ in the framework of
 FRW cosmology. The corresponding energy density of ECHDE is
 given by
 \be{eq7}
 \rho_\Lambda=3n^2 m_p^2 H^2+
 \alpha H^4 \ln\left(\frac{m_p^2}{H^2}\right)+\beta H^4.
 \ee
 Since (EC)HDE is the vacuum fluctuation energy, one can image
 that the universe is dominated by ECHDE in the very early
 stage. The Friedmann equation reads $3m_p^2 H^2=\rho_\Lambda$.
 For convenience, we introduce a new dimensionless quantity
 \be{eq8}
 x\equiv\frac{H^2}{m_p^2}\geq 0\,.
 \ee
 So, the Friedmann equation can be recast as
 \be{eq9}
 n^2+\frac{1}{3}x\left(-\alpha\ln x+\beta\right)=1.
 \ee
 For constant $n$, $\alpha$ and $\beta$, we can expect that
 $x$ is also constant. For the case of $\alpha=0$ but
 $\beta\not=0$, it is easy to find that
 \be{eq10}
 x=3\left(1-n^2\right)\beta^{-1}=const..
 \ee
 Noting that $x\equiv H^2/m_p^2\ge 0$ by definition, for the
 case of $\alpha=0$ but $\beta\not=0$, we should require
 $n\le 1$ and $n\ge 1$ for $\beta>0$ and $\beta<0$,
 respectively. On the other hand, for the case of
 $\alpha\not=0$, we can solve Eq.~(\ref{eq9}) and find that
 \be{eq11}
 x=\frac{3}{\alpha}\left(n^2-1\right)\left\{
 {\rm ProductLog}\left[\frac{3}{\alpha}\left(n^2-1\right)
 e^{-\beta/\alpha}\right]\right\}^{-1}=const.,
 \ee
 where the special function ${\rm ProductLog}\,[z]$ gives the
 principal solution for $w$ in $z=we^w$. In Table~\ref{tab1},
 for examples, we present some numerical solutions of $x$ for
 various given $n$, $\alpha$ and $\beta$.

Note that ECHDE reduces to the ordinary HDE when $\alpha=\beta=0$,
 or, speaking strictly $x\left(-\alpha\ln x+\beta\right)$ can
 be ignored. In this case, from Eq.~(\ref{eq9}) we get $n^2=1$.
 This is not surprising. In~\cite{r9,r10,r11}, it is found that
 the EoS of HDE $w_\Lambda=-1/3-2\sqrt{\Omega_\Lambda}/(3n)$ in
 the framework of FRW cosmology. Only if $n=1$, $w_\Lambda$
 can be $-1$ when $\Omega_\Lambda=1$. So, if $n\not=1$ and
 $\alpha=\beta=0$ [or speaking strictly
 $x\left(-\alpha\ln x+\beta\right)$ can be ignored],
 $L=H^{-1}=const.$ is {\em unphysical}.

In fact, $x=const.$ means $H=const.$ which corresponds to an
 inflation phase for $H>0$ or an exponential collapse phase
 for $H<0$. It is easy to find that in this phase
 $\rho_\Lambda=const.$ and the EoS of ECHDE $w_\Lambda=-1$.
 From Eq.~(\ref{eq10}) and Table~\ref{tab1}, we see that
 non-zero $\alpha$ and/or $\beta$ can help to achieve the
 inflation phase {\em without} requiring $n=1$.

%==================== table 1 ====================
 \begin{table}[tbp]
 \begin{center}
 \begin{tabular}{c|c|c||c} \hline\hline
 $n$ & $\ \alpha\ $ & $\ \beta\ $ & $\ x\ $ \\ \hline
 ~~~1.1~~~ & ~~~~~~1/2~~~~~~ & ~~~~~~0~~~~~~ & ~~~1.92456~~~ \\
 1.1 & $-1/2$ & $-2/3$ & 2.05457 \\
 1.1 & 1/2 & 1/3 & 2.9749 \\
 1.1 & $-1/2$ & $-3/2$ & 18.7823 \\
 1.1 & 1/2 & 3/2 & 21.309 \\
 0.9 & $-1/2$ & 0 & 1.85118 \\
 0.9 & 1/2 & 2/3 & 2.32176 \\
 0.9 & $-1/2$ & $-1/3$ & 2.88974 \\
 0.9 & 1/2 & 3/2 & 18.9105 \\
 0.9 & $-1/2$ & $-3/2$ & 21.1954 \\
 \hline\hline
 \end{tabular}
 \end{center}
 \caption{\label{tab1} Some numerical solutions of $x$ from
 Eq.~(\ref{eq9}), for various given $n$, $\alpha$ and $\beta$.}
 \end{table}
%=================================================

%============================= section 2.2 ===================================

\subsection{The case in loop quantum cosmology}\label{sec2b}
Next, we consider the ECHDE with $L=H^{-1}$ in the framework of
 loop quantum cosmology (LQC). In LQC, the Friedmann equation
 has been modified to~\cite{r34,r35,r36,r37,r38,r39,r40,r41,r42}
 \be{eq12}
 H^2=\frac{\rho_{tot}}{3m_p^2}\left(1-
 \frac{\rho_{tot}}{\rho_c}\right),
 \ee
 where $\rho_{tot}$ is the total energy density, and
 \be{eq13}
 \rho_c=4\sqrt{3}\gamma^{-3}m_p^4\,,
 \ee
 in which $\gamma$ is the dimensionless Barbero-Immirzi
 parameter (it is suggested that $\gamma\simeq 0.2375$ by the
 black hole thermodynamics in LQG~\cite{r43}). Similar to the
 previous subsection, we consider the universe which is
 dominated by ECHDE in the very early stage. In this case,
 $\rho_{tot}=\rho_\Lambda$. The Friedmann equation~(\ref{eq12})
 can be recast as
 \be{eq14}
 \left[n^2+\frac{1}{3}x\left(-\alpha\ln x+\beta\right)
 \right]\times\left\{1-\frac{m_p^4}{\rho_c}\,x\left[
 3n^2+x\left(-\alpha\ln x+\beta\right)\right]\right\}=1.
 \ee
 For constant $n$, $\alpha$ and $\beta$, we can expect that
 $x$ is also constant. It is difficult to get analytical
 solution of $x$ from Eq.~(\ref{eq14}). We instead present some
 numerical solutions of $x$ for various given $n$, $\alpha$ and
 $\beta$ in Table~\ref{tab2} for examples.

Note that ECHDE reduces to ordinary HDE when $\alpha=\beta=0$,
 or, speaking strictly $x\left(-\alpha\ln x+\beta\right)$ can
 be ignored. In this case, from Eq.~(\ref{eq14}) we get
 \be{eq15}
 x=\frac{\left(n^2-1\right)\rho_c}{3n^4 m_p^4}=const.
 \ee
 Noting that $x\equiv H^2/m_p^2\ge 0$ by definition, for this
 case, we should require $n\ge 1$. So, if $n<1$ and
 $\alpha=\beta=0$ [or speaking strictly
 $x\left(-\alpha\ln x+\beta\right)$ can be ignored],
 $L=H^{-1}=const.$ is {\em unphysical}.

In fact, $x=const.$ means $H=const.$ which corresponds to an
 inflation phase for $H>0$ or an exponential collapse phase
 for $H<0$. It is easy to find that in this phase
 $\rho_\Lambda=const.$ and the EoS of ECHDE $w_\Lambda=-1$.
 From Table~\ref{tab2}, we see that
 non-zero $\alpha$ and/or $\beta$ can help to achieve the
 inflation phase {\em without} requiring $n\ge 1$.

%==================== table 2 ====================
 \begin{table}[tbp]
 \begin{center}
 \begin{tabular}{c|c|c||c} \hline\hline
 $n$ & $\ \alpha\ $ & $\ \beta\ $ & $\ x\ $ \\ \hline
 ~~~0.9~~~ & ~~~~$-1/2$~~~~ & ~~~~~~0~~~~~~ & ~~~1.89258~~~ \\
 0.9 & 0 & 1/3 & 1.80633 \\
 0.9 & $-1/2$ & $-1/3$ & 2.96571 \\
 0.9 & 1/2 & 3/4 & 2.92942 \\
 0.9 & $-1/2$ & $-1/2$ & 20.0024 \\
 0.9 & 1/2 & 3/2 & 18.0286 \\
 1.1 & 1/2 & 0 & 1.88378 \\
 1.1 & 0 & $-1/3$ & 1.79432 \\
 1.1 & $-1/2$ & $-2/3$ & 2.2405 \\
 1.1 & 1/2 & 1/3 & 2.90083 \\
 1.1 & $-1/2$ & $-1/2$ & 19.5806 \\
 1.1 & 1/2 & 3/2 & 20.3902 \\
 \hline\hline
 \end{tabular}
 \end{center}
 \caption{\label{tab2} Some numerical solutions of $x$ from
 Eq.~(\ref{eq14}), for various given $n$, $\alpha$ and $\beta$.}
 \end{table}
%=================================================

%============================= section 3 ===================================

\section{Inflation phase driven by ECHDE with $L=R_h$}\label{sec3}
In this section, we consider the universe which is dominated
 by ECHDE with $L=R_h$ at the very early stage. Similar to the
 previous section, we also work in both frameworks of FRW
 cosmology and loop quantum cosmology (LQC). Before plunging
 into particular cases, let us start from the very beginning.
 Differentiating Eq.~(\ref{eq6}), we obtain
 \bea
 \dot{\rho}_\Lambda&=&(-4)\frac{\dot{L}}{L}\left[
 \frac{3}{2}n^2 m_p^2 L^{-2}+\alpha L^{-4}\ln\left(m_p^2 L^2\right)
 -\frac{\alpha}{2}L^{-4}+\beta L^{-4}\right]\label{eq16}\\
 &=&(-4)\frac{\dot{L}}{L}\left(\rho_\Lambda-
 \frac{3}{2}n^2 m_p^2 L^{-2}-
 \frac{\alpha}{2}L^{-4}\right).\label{eq17}
 \eea
 Note that these results have nothing to do with the particular
 choice of $L$ and the Friedmann equation. Then, we consider
 the case with $L=R_h$. Differentiating Eq.~(\ref{eq4}), we
 have
 \be{eq18}
 \frac{\dot{L}}{L}=H-\frac{1}{L}\,.
 \ee

%============================= section 3.1 ===================================

\subsection{The case in FRW cosmology}\label{sec3a}
Now, we consider the case in FRW cosmology first. The corresponding
 Friedmann equation reads $3m_p^2 H^2=\rho_\Lambda$. Substituting
 into Eq.~(\ref{eq17}) and using Eq.~(\ref{eq18}), we get
 \be{eq19}
 6m_p^2 H\dot{H}=(-4)\left(H-\frac{1}{L}\right)
 \left(3m_p^2 H^2-\frac{3}{2}n^2 m_p^2 L^{-2}-
 \frac{\alpha}{2}L^{-4}\right).
 \ee
 In fact, it is very difficult to solve this equation, because
 $L=R_h$ is given in the form of an integration, namely
 Eq.~(\ref{eq4}). By observation, one can see that
 $L=H^{-1}=const.$ is a special solution to Eq.~(\ref{eq19})
 combining with $\dot{H}=\dot{\rho}_\Lambda=0$ and
 $\disp R_h=a\int_a^\infty\frac{d\tilde{a}}{H\tilde{a}^2}=H^{-1}$.
 However, since $R_h$ and $H$ evolve separately, $R_h=H^{-1}$
 holds only when $H=const.$; in the other cases, $R_h\not=H^{-1}$.
 Seemingly, there is no reason to say that the inflation solution
 $L=R_h=H^{-1}=const.$ is necessary.

Let us change to another perspective. In the following, the
 method of dynamical system will play a key role. In the
 literature (see e.g.~\cite{r44,r45,r46,r47,r48}),
 the method of dynamical system has been extensively used to
 alleviate the cosmological coincidence problem ``why the
 densities of dark energy and pressureless matter are
 comparable recently?''. See e.g.~\cite{r49} for some
 reviews on dynamical system.

Using the relation $f^\prime=\dot{f}/H$ for any function
 $f$ (here a prime denotes the derivative with respect to the
 so-called $e$-folding time $N\equiv\ln a$), we can recast
 Eqs.~(\ref{eq19}) and (\ref{eq18}) as
 \bea
 &&H^\prime=\frac{-2}{3m_p^2 H^2}\left(H-\frac{1}{L}\right)
 \left(3m_p^2 H^2-\frac{3}{2}n^2 m_p^2 L^{-2}-
 \frac{\alpha}{2}L^{-4}\right),\label{eq20}\\
 &&L^\prime=L-H^{-1}.\label{eq21}
 \eea
 Obviously, Eqs.~(\ref{eq20}) and (\ref{eq21}) form an
 autonomous system of $H$ and $L$. The fixed points of this
 autonomous system can be determined by imposing
 $H^\prime=L^\prime=0$. It is easy to see that the only
 fixed point is given by $L=H^{-1}=const.$. Although $L=R_h$
 and $H$ evolve separately, for this dynamical system, the
 universe will enter the attractor $L=R_h=H^{-1}=const.$
 sooner or later, regardless of the initial
 conditions~\cite{r49}. Over there, all the following things
 are the same as in Sec.~\ref{sec2a}. The universe undergoes
 an inflation phase for $H>0$ or an exponential collapse phase
 for $H<0$.

%============================= section 3.2 ===================================

\subsection{The case in loop quantum cosmology}\label{sec3b}
As mentioned above, in LQC the Friedmann equation has been
 modified to Eq.~(\ref{eq12}) with $\rho_{tot}=\rho_\Lambda$.
 So, we get
 \be{eq22}
 H^\prime=\frac{\dot{\rho}_\Lambda}{6m_p^2 H^2}
 \left(1-\frac{2\rho_\Lambda}{\rho_c}\right).
 \ee
 Using Eqs.~(\ref{eq16}), (\ref{eq18}) and (\ref{eq6}), we can
 finally obtain
 \be{eq23}
 H^\prime=\left(H-\frac{1}{L}\right)F(L)\,,
 \ee
 where $F(L)$ is a very involved function of $L$ and hence we
 do not present its explicit expression here. Again,
 Eqs.~(\ref{eq23}) and (\ref{eq21}) form an autonomous system
 of $H$ and $L$. Obviously, the only fixed point is also given
 by $L=H^{-1}=const.$. Although $L=R_h$ and $H$ evolve
 separately, for this dynamical system, the universe will
 enter the attractor $L=R_h=H^{-1}=const.$ sooner or later,
 regardless of the initial conditions~\cite{r49}. Over there,
 all the following things are the same as in Sec.~\ref{sec2b}.
 The universe undergoes an inflation phase for $H>0$ or an
 exponential collapse phase for $H<0$.

%============================= section 4 ===================================

\section{Evolution history of the universe with ECHDE}\label{sec4}
Now, we consider the entire history of the universe with ECHDE.
 At the very early stage, the universe is dominated by ECHDE
 with $L=H^{-1}$ or $L=R_h$. We assume that the universe expands
 initially and hence $H>0$ at the beginning. For the case with
 $L=H^{-1}$, as shown in Sec.~\ref{sec2}, the universe
 undergoes an inflation phase naturally in both frameworks of
 FRW cosmology and LQC. For the case with $L=R_h$, the situation
 is more complicated. As shown in Sec.~\ref{sec3}, the universe
 enters an inflation phase sooner or later in both frameworks of
 FRW cosmology and LQC. This is similar to the issue of
 inflation attractor studied in the literature
 (e.g.~\cite{r50,r51,r52}). We assume that at the end of
 inflation phase ECHDE decays into radiation and matter; this
 might be achieved through the mechanism similar to the one of
 reheating or preheating~\cite{r53}. Then, the universe
 undergoes the familiar radiation-dominated and
 matter-dominated epochs. In these two  epochs, since the
 universe is much larger, the entropy-corrected terms to ECHDE,
 namely the last two terms in Eq.~(\ref{eq6}), can be safely
 ignored. On the other hand, since the total energy density
 becomes smaller, the correction to the Friedmann equation in
 LQC can also be safely ignored. Therefore, the evolutions of
 the universe in the radiation-dominated and matter-dominated
 epochs become the same as the ones with ordinary HDE in the
 framework of FRW cosmology. For HDE with $L=H^{-1}$, as shown
 in~\cite{r8,r9}, however, in this case the resulting EoS is
 equal to zero, which cannot accelerate the expansion of the
 universe. This is contrary to the cosmological observations.
 So, ECHDE with $H^{-1}$ is no longer of interest. It is
 worth noting that Sec.~\ref{sec2} is still worthwhile, since
 ECHDE with $L=R_h$ has the same results as in Sec.~\ref{sec2}
 when the universe eventually enters the inflation attractor
 $L=R_h=H^{-1}=const.$. We stress that $L=R_h=H^{-1}=const.$
 only holds in the inflation phase. In the radiation-dominated
 and matter-dominated epochs, $L=R_h$ and $H$ evolve
 separately. For ECHDE with $L=R_h$, there are three cases with
 $n>1$, $n=1$ and $n<1$. We will discuss them one by one in
 the following.

As mentioned above, in the radiation-dominated and
 matter-dominated epochs, ECHDE with $L=R_h$ and any $n$ in
 both frameworks of FRW cosmology and LQC reduces to the
 ordinary HDE with $L=R_h$ and corresponding $n$ in FRW
 cosmology, since the entropy-corrected terms to ECHDE and the
 correction to Friedmann equation in LQC can be safely ignored.
 So, their results are the same as the ones obtained
 in~\cite{r9,r10,r11}. In particular, its EoS is given by
 $w_\Lambda=-1/3-2\sqrt{\Omega_\Lambda}/(3n)$.
 Since $w_\Lambda<-1/3$, HDE will dominate the universe in the
 late time. For the case with $n\ge 1$, we have
 $w_\Lambda\ge -1$ always. Thus, the density of HDE decreases
 always, and hence the correction to Friedmann equation in LQC
 can also be ignored always. For the case with $n>1$, the
 universe undergoes accelerated (but not exponential) expansion
 forever. Here, one might argue that from the results in
 sections~\ref{sec2a} and \ref{sec3a} the universe will enter
 the $L=H^{-1}=const.$ phase again. This is a delusion in
 fact. Noting that in this case the entropy-corrected terms
 $x\left(-\alpha\ln x+\beta\right)$ can be ignored because
 the universe is vary large, as shown in the end of
 Sec.~\ref{sec2a}, $L=H^{-1}=const.$ is {\em unphysical} for
 $n\not=1$. On the other hand, for the case with $n=1$, the
 universe can enter a de Sitter phase eventually.

For the case with $n<1$, the situation is more complicated. In
 this case, $w_\Lambda$ can cross the phantom divide, as shown
 in e.g.~\cite{r11}. $w_\Lambda<-1$ at the late time and HDE
 becomes phantom-like. Thus, the density of HDE increases
 instead. In the framework of FRW cosmology, the universe will
 end in a big rip. It is worth noting that if $n<1$ and the
 entropy-corrected terms $x\left(-\alpha\ln x+\beta\right)$
 can be ignored because the universe is vary large, as shown
 in the end of Sec.~\ref{sec2b}, $L=H^{-1}=const.$ is
 {\em unphysical}. Therefore, the universe cannot enter
 the $L=H^{-1}=const.$ (de Sitter) phase again in both
 frameworks of FRW cosmology and LQC. On the other hand, in the
 framework of LQC, the correction to the Friedmann equation
 cannot be ignored sooner or later, because $w_\Lambda<-1$ and
 hence the density of HDE increases continuously. Similar to
 the phantom bounce (speaking strictly, turnaround) in the
 framework of LQC studied in~\cite{r54,r55}, the universe
 reaches a maximum scale factor $a_{max}$ when
 $\rho_\Lambda=\rho_c$, and then the bounce (turnaround)
 happens. The universe contracts with $H<0$. The density of HDE
 decreases, whereas the densities of matter and radiation
 increase. Once again, the universe undergoes matter-dominated
 and radiation-dominated epochs sequently. We assume that ECHDE
 decays into radiation and matter at the end of the inflation
 phase in the early stage according to the decay rate
 $\Gamma\propto H$. Now, $H$ becomes negative, and hence the
 procedure inverts. Radiation and matter decay into ECHDE now.
 Eventually, ECHDE dominates again with $H<0$. According to
 sections~\ref{sec2} and \ref{sec3}, the universe will collapse
 exponentially and end in a singularity. We hope that the
 unknown trans-Planckian physics can prevent it, and even make
 the universe bounce again (in this case the cyclic universe
 is possible).

%============================= section 5 ===================================

\section{Entropy-corrected agegraphic dark energy}\label{sec5}
Following the line of quantum fluctuations of spacetime,
 in Refs.~\cite{r56,r57,r58}, by using the so-called
 K\'{a}rolyh\'{a}zy relation
 $\delta\tau=\lambda t_p^{2/3}\tau^{1/3}$~\cite{r59} and
 the well-known time-energy uncertainty relation
 $E_{\delta\tau^3}\sim\tau^{-1}$, it was argued
 that the energy density of metric fluctuations of
 Minkowski spacetime is given by
 \be{eq24}
 \rho_\Lambda\sim\frac{E_{\delta\tau^3}}{\delta\tau^3}\sim
 \frac{1}{t_p^2 \tau^2}\sim\frac{m_p^2}{\tau^2}\,,
 \ee
 where $l_p=t_p=1/m_p$ with $l_p$, $t_p$ and $m_p$ being
 the reduced Planck length, time and mass, respectively.
 In~\cite{r58}, as the most natural choice, the time scale
 $\tau$ in Eq.~(\ref{eq24}) was chosen to be the age of the
 universe
 \be{eq25}
 T=\int_0^a\frac{d\tilde{a}}{H\tilde{a}}\,,
 \ee
 and the density of this so-called ``agegraphic dark energy''
 (ADE) reads
 \be{eq26}
 \rho_\Lambda=3n^2m_p^2 T^{-2}.
 \ee
 We refer to e.g.~\cite{r58,r60,r61,r62,r63} for the works on
 ADE. However, the ADE model might contain an
 inconsistency~\cite{r31}. To overcome this problem, the new
 agegraphic dark energy (NADE) was proposed in ~\cite{r31,r32}.
 In the NADE model, the time scale $\tau$ in Eq.~(\ref{eq24})
 was chosen to be the conformal time
 \be{eq27}
 \eta\equiv\int_0^a\frac{d\tilde{a}}{H\tilde{a}^2}\,.
 \ee
 The density of NADE is given by
 \be{eq28}
 \rho_\Lambda=3n^2m_p^2\eta^{-2}.
 \ee
 We refer to e.g.~\cite{r31,r32,r33,r64,r65} for the works on NADE.

Here, we would like to consider the so-called
 ``entropy-corrected agegraphic dark energy'' (ECADE) whose
 $L$ in Eq.~(\ref{eq6}) is chosen to be the conformal time
 $\eta$ (note that we set units $\hbar=c=1$ throughout, one
 can use the terms like length and time interchangeably). In
 fact, ECADE is the entropy-corrected version of
 NADE~\cite{r31,r32}. The density of ECADE is given by
 \be{eq29}
 \rho_\Lambda=3n^2 m_p^2 \eta^{-2}+
 \alpha \eta^{-4}\ln\left(m_p^2 \eta^2\right)+\beta \eta^{-4}.
 \ee
 In~\cite{r66}, it is argued that HDE and (N)ADE might share
 the same origin. We agree with this point of view. As an
 additional example, we refer to~\cite{r33} in which NADE with
 generalized uncertainty principle (GUP) was studied. The
 corresponding density of NADE with GUP is derived
 in~\cite{r33}, namely
 \be{eq30}
 \rho_\Lambda=3n^2 m_p^2 \eta^{-2}+\beta \eta^{-4}.
 \ee
 It is easy to see that NADE with GUP can be regarded as a
 special case of ECADE with $\alpha=0$.

\newpage % used here just for a more comfortable typesetting

For ECADE with $L=\eta$, similar to Sec.~\ref{sec3}, we find
 that $\dot{L}/L=(La)^{-1}$ and $L^\prime=(Ha)^{-1}$. The scale
 factor $a$ appears explicitly. So, we cannot form an
 autonomous system of $H$ and $L$. There is no inflation
 attractor in ECADE model in the very early stage. On the other
 hand, ECADE reduces to NADE when the universe is large. As
 shown in~\cite{r31,r32}, the EoS of NADE reads
 $w_\Lambda=-1+2\sqrt{\Omega_\Lambda}/(3na)$, which is lager
 than $-1$ always. So, $\rho_\Lambda$ of NADE decreases always,
 and hence the correction to Friedmann equation in LQC can also
 be ignored always. Therefore, there is no bounce (or
 turnaround) in the ECADE model, regardless the value of $n$.
 All these results make the ECADE model not so attractive.

%============================= section 6 ===================================

\section{Final remarks}\label{sec6}
In this work, we have considered many aspects of ECHDE. There
 are some remarks further. Firstly, the mechanism for ECHDE
 decaying into radiation and matter should be studied in
 details. We leave it to the future work. Secondly, the
 primordial scalar power spectrum of inflation with HDE was
 calculated in~\cite{r67}. However, the inflation studied
 in~\cite{r67} was driven by a normal slow-roll scalar field,
 while HDE merely gives small corrections to the primordial
 scalar power spectrum of this scalar inflaton. However, in
 the ECHDE model, the inflation was driven by ECHDE
 {\em itself}. This makes things different. Thirdly, we propose
 that the curvature perturbation in the ECHDE model might be
 generated through the curvaton~\cite{r68,r69}. In the curvaton
 scenario, the generation of curvature perturbation by the
 curvaton requires no assumption about the nature of inflation
 beyond the requirement that $H\simeq const.$~\cite{r68}. So,
 it is very suitable to the inflation driven by ECHDE. We also
 leave the detailed calculation of the primordial power
 spectrum to the future work. Fourthly, in the literature
 (see e.g.~\cite{r20,r29,r30} for some brief reviews),
 one can also consider a more general entropy-area relation
 \be{eq31}
 S_{BH}=\frac{A}{4G}+\tilde{\alpha}_1\ln\frac{A}{4G}
 +\tilde{\alpha}_2\frac{4G}{A}+\tilde{\alpha}_3\,,
 \ee
 where $\tilde{\alpha}_i$ are dimensionless constants of order
 unity. Correspondingly, the density of ECHDE should be
 \be{eq32}
 \rho_\Lambda=3n^2 m_p^2 L^{-2}+
 \alpha_1 L^{-4}\ln\left(m_p^2 L^2\right)+
 \alpha_2 m_p^{-2} L^{-6}+\alpha_3 L^{-4}.
 \ee
 Finally, the relation between
 ECHDE and NADE with GUP might hint some deep properties of
 quantum gravity. This also deserves further investigation.

%============================= acknowledgements ===================================

\section*{ACKNOWLEDGEMENTS}
We are grateful to Professors Rong-Gen~Cai, Shuang-Nan~Zhang,
 and Miao~Li for helpful discussions. We also thank Minzi~Feng,
 as well as Xin~Zhang, Chang-Jun~Gao and Pu-Xun~Wu, for kind
 help and discussions. This work was supported by the
 Excellent Young Scholars Research Fund of Beijing Institute
 of Technology.

 \renewcommand{\baselinestretch}{1.1}

%============================= references ===================================

\end{document}